\documentclass[aip,amsmath,amssymb,reprint,floatfix]{revtex4-1}
\usepackage{graphicx}
\usepackage{xcolor}
\usepackage{dcolumn}
\usepackage{bm}
\usepackage{natbib}
\usepackage[utf8]{inputenc}
\usepackage[T1]{fontenc}
\usepackage{mathptmx}
\usepackage[colorlinks, citecolor=red]{hyperref}

\begin{document}
\title{Vanadium gate-controlled Josephson half-wave nanorectifier}
\author{C. Puglia}
\email{claudio.puglia@df.unipi.it}
\affiliation{Dipartimento di Fisica, Università di Pisa, Largo Bruno Pontecorvo 3, I-56127 Pisa, Italy.}
\affiliation{NEST, Istituto Nanoscienze-CNR and Scuola Normale Superiore, I-56127 Pisa, Italy.}
\author{G. De Simoni}
\email{giorgio.desimoni@nano.cnr.it}
\affiliation{NEST, Istituto Nanoscienze-CNR and Scuola Normale Superiore, I-56127 Pisa, Italy.}
\author{N. Ligato}
\affiliation{NEST, Istituto Nanoscienze-CNR and Scuola Normale Superiore, I-56127 Pisa, Italy.}
\author{F. Giazotto}
\email{francesco.giazotto@sns.it}
\affiliation{NEST, Istituto Nanoscienze-CNR and Scuola Normale Superiore, I-56127 Pisa, Italy.}
\begin{abstract}
Recently, the possibility to tune the critical current of conventional metallic superconductors via electrostatic gating was shown in wires, Josephson weak-links and superconductor-normal metal-superconductor junctions. Here we exploit such a technique to demonstrate a gate-controlled vanadium-based Dayem nano-bridge operated as a \emph{half-wave} rectifier at $3$ K. Our devices exploit the gate-driven modulation of the critical current of the Josephson junction, and the resulting steep variation of its normal-state resistance, to convert an AC signal applied to the gate electrode into a DC one across the junction. All-metallic superconducting  gated rectifiers could provide the enabling technology to realize tunable photon detectors and diodes useful for superconducting electronics circuitry.
\end{abstract}
\maketitle

In the last few years, the possibility to suppress the critical current in conventional Bardeen-Cooper-Schrieffer (BCS) superconducting wires
\cite{DeSimoni2018}, Dayem bridge (DB) Josephson junctions (JJs) \cite{Paolucci2018,Paolucci2019,Bours2020,DeSimoni2020} and superconductor-normal metal-superconductor weak-links \cite{DeSimoni2019} via electrostatic gating was demonstrated \cite{Paolucci2019a}. 
Such a technique is at the basis of a class of innovative all-metallic superconducting Josephson transistors, where a gate electrode is used to modulate down to full suppression the critical current of a superconducting channel, in analogy to what occurs for the current in standard gated semiconductor transistors. 
Furthermore, the impact of the electrostatic gating on the macroscopic phase, and on the phase slip dynamics was proved in superconducting quantum interference devices (SQUIDs) \cite{Paolucci2019b} and in gated titanium DBs, respectively \cite{Puglia2020}. 
So far, the microscopic origin of the gating effect in  metallic superconductors is still unclear, and the debate in the search of a satisfactory explanation for the aforementioned phenomenology has just started. \cite{Virtanen2019,Mercaldo2019,Bours2020}. 
Yet, several technological implementations based on gated superconducting  transistors for both classical and quantum computation architectures were proposed \cite{Paolucci2019a} til now, and promise new paradigms to realize innovative superconducting  circuits. 
These are expected to benefit from operation frequencies close to those typical of rapid single flux quantum devices (up to $\sim750$ GHz) \cite{Chen1998}, but in a \emph{voltage-driven} fashion which would provide a direct and convenient interface with complementary metal-oxide technology (CMOS).

\begin{figure}[t!]
\includegraphics[width=\linewidth]{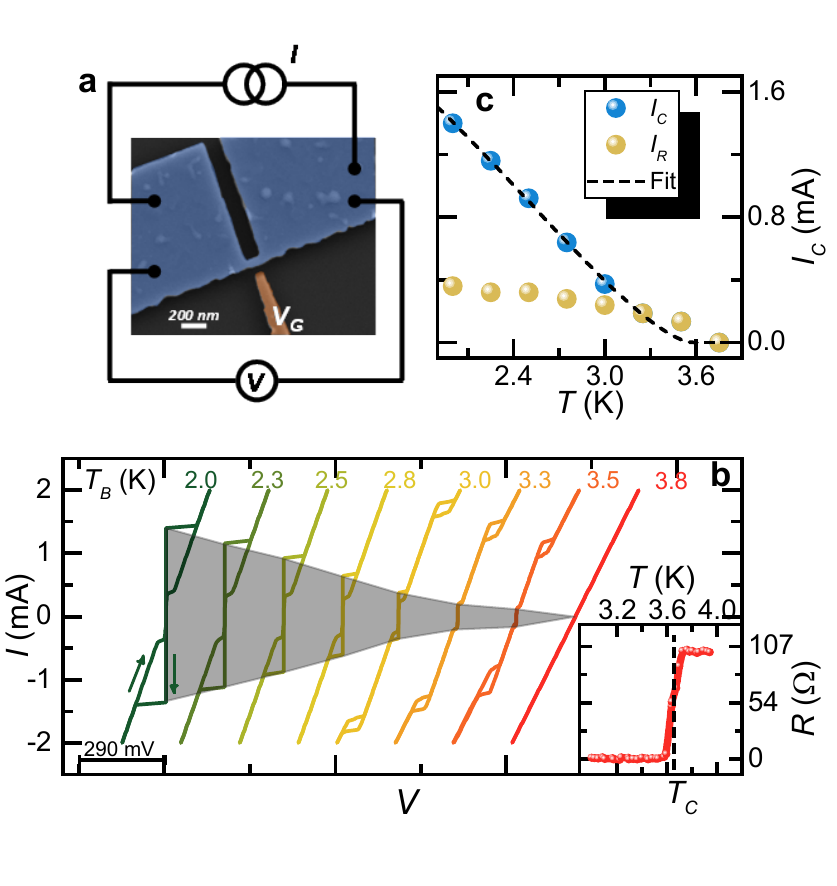}
\caption{\label{fig:fig1} (a) False-color scanning electron micrograph of a typical gate-controlled vanadium  JJ nanotransistor. The DB  constriction (light blue) is current biased, and the voltage drop is measured in a 4-wire configuration. The electrostatic field is applied via the gate electrode (orange). (b) Back and forth current $I$ vs voltage $V$ characteristics of a representative device measured at different bath temperatures from 2.0 K to 3.8 K. The curves are horizontally offset for clarity. The inset shows the 4-wire lock-in measurement (with biasing current $I\simeq 15$ nA) of the $R$ vs $T$ characteristic of the junction. The superconducting transition occurs at $T_C\simeq3.6K$. (c) Evolution of the critical current $I_C$ (light blue dots) and retrapping current $I_R$ (yellow dots) vs bath temperature $T$. The black dashed line represents the best fit of $I_C (T)$ with the   Bardeen's theory, which yields $I_C^0=(2.2\pm0.1)$ mA and $T_C^{fit}=(3.62\pm0.07)$ K as fitting parameters.}
\end{figure}
Here, we demonstrate a vanadium (V) Josephson half-wave rectifier based on a scheme recently proposed \cite{DeSimoni2020} which exploits the strong non-linearity of the resistance ($R$) of a gated superconducting nanotransistor  \textit{vs} gate voltage ($V_G$). 
Such a device might form the rectifying core of a photon sensor operating at high frequencies, relevant, for instance, for cosmological applications \cite{Rowan-Robinson2009}, or could be exploited as a comparator \cite{Mukherjee2018} to be used in all-superconducting digital electronic circuits\cite{Paolucci2019a}. 
The potentially-high operation frequency, the ease of fabrication and the simple geometry of our nanodevices makes them  attractive as an effective alternative to conventional all-semiconductor \cite{Valizadeh2016} and semiconductor-superconductor hybrid \cite{Nishino1989,Clark1980,Gray1978} field -effect-transistor-based technologies. \cite{Deware2017,AnuarRosli2015} 
Below, we focus on the manipulation of the dissipative response of the V-based Josephson weak-link via the application of a voltage to a gate capacitively-coupled to the junction, and show  its time-resolved response to periodic square-wave and sinusoidal signals applied to the gate thereby leading to electric rectification.

Our device consists of a planar 60-nm-tick, 160-nm-long, 90-nm-wide V Dayem bridge JJ with a 70-nm-apart, 120-nm-wide side-gate aligned to the constriction. The device fabrication was performed on a silicon oxide substrate with a single-step electron beam lithography 
followed by vanadium deposition performed at a rate of 0.36 nm/s
in an ultra-high vacuum electron-beam evaporator with a base pressure of $\sim 10^{-11}$ Torr \cite{garcia2009josephson,spathis2011hybrid,quaranta2011cooling,giazotto2011josephson,ronzani2013micro,ronzani2014balanced,ligato2017high}. 
Figure \ref{fig:fig1}(a) shows the false color scanning electron micrograph of a representative V JJ transistor along with the 4-wire biasing scheme used for the low-temperature current vs voltage ($I-V$) characterization of the devices performed in a filtered cryogen-free $^3$He-$^4$He dilution refrigerator. 

For temperatures below $\sim 3.5$ K, the V JJ is superconducting and exhibit, at $2$ K, a critical current $I_C$ of $\sim 1.42$ mA, a retrapping current $I_R\sim 380\ \mu$A, and a normal-state resistance $R_N$ of around $\sim 110\ \Omega$. 
The forward and backward current $I$ vs $V$ characteristics for temperatures $T$ ranging from 2.0 K to 3.8 K are shown in Fig. \ref{fig:fig1}(b), where the full evolution of $I_C$ as a function of $T$ is highlighted by the grey area. 
The hysteretic behaviour of the curves stems from quasiparticle over-heating when the junction switches from the superconducting  to the normal state \cite{Skocpol1974,courtois2008origin}. Moreover, the additional features present on the $I-V$ characteristics at higher current are likely to be related to the switching of the DB banks to the normal state.
The decay of both $I_S$ and $I_R$ with the  temperature is shown in Fig. \ref{fig:fig1}(c), where the black dashed line represents a fit of the $I_C$ vs $T$ characteristic performed with the Bardeen's equation:
$I_C(T)=I_C^0\left[1-\left(\frac{T}{T_C}\right)^2\right]^{\frac{3}{2}}$,
where $I_C^0$ is the zero-temperature critical current, and $T_C$ is the DB critical temperature.
The fitting procedure yielded $I_C^0=(2.2\pm0.1)$ mA and $T_C^{fit}=(3.62\pm 0.07)$ K. The latter value is in agreement with the experimental critical temperature determined from the 4-wire lock-in measurement of the junction resistance vs bath temperature $T$, as shown in the inset of Fig. \ref{fig:fig1}(b). 
\begin{figure}[ht]
\includegraphics[width=\linewidth]{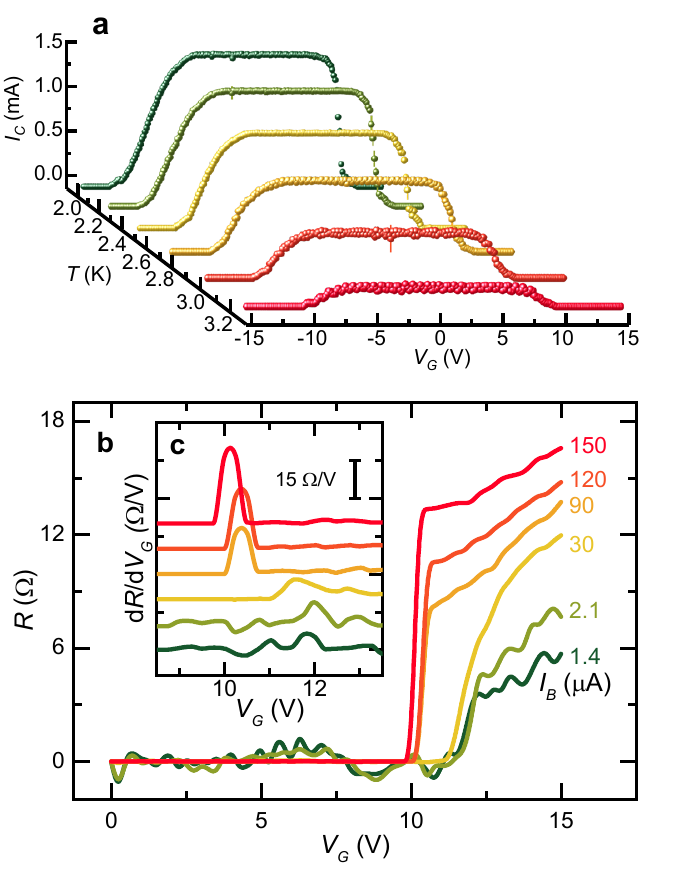}
\caption{\label{fig:fig2} 
(a) $I_C$ vs $V_G$ characteristics for different temperatures ranging from 2.0 to 3.3 K. Data were computed averaging 25 acquisitions of the switching current. The error bars represent the standard deviation of the ensemble of measurements. 
(b)  Resistance $R=V/I_B$ vs gate voltage $V_G$ for different values of bias current $I_B$ from 1.4 to 150 $\mu$A. 
(c) Transresistance $\frac{dR}{dV_G}$ vs gate voltage $V_G$ for the same bias currents of panel (b).  Measurements of panels (b) and (c) were carried out at $T_B=3.0$ K.}
\end{figure}

The preliminary characterization of the gating effect on $I_C$ and $R$ was carried out by acquiring the forward and backward $I-V$ characteristics of a representative vanadium DB as function of the voltage applied to the gate electrode, $V_G$. 
Figure \ref{fig:fig2}(a) displays the $I_C(V_G)$ characteristics extracted from the $I-V$ acquired for several bath temperatures. The modulation of $I_C$ is almost symmetric for $V_G\longrightarrow -V_G$, and the gate effect persists up to $\sim3.3$ K, i. e., $\sim0.9 T_C$. 
We  note that a sharper suppression of $I_C$ for positive values of $V_G$ seems to be in contrast to a possible field emitted electrons-driven effect, since  electrons extraction from the gate (occurring at negative gate bias) is, in principle, more effective than that from the DB (occurring at positive gate bias) owing to the substantial asymmetry of the device geometry.
\begin{figure*}[htb]
\includegraphics[width=\textwidth]{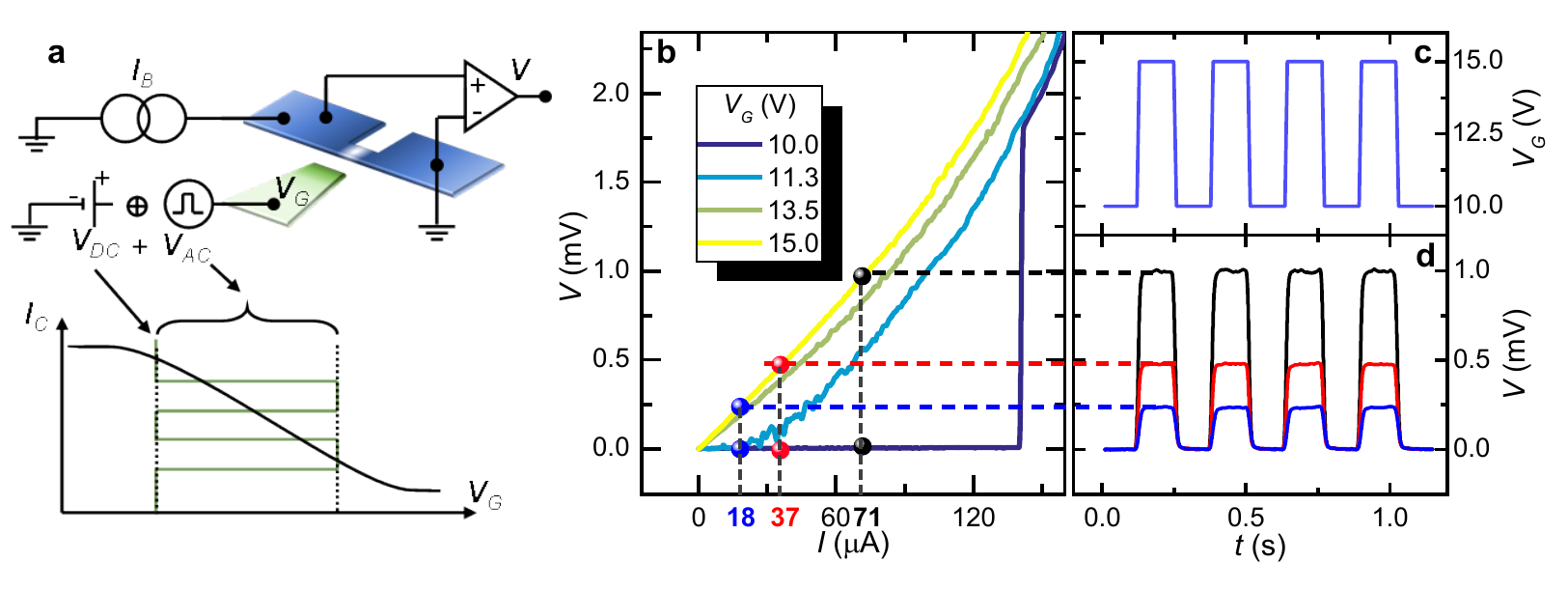}
\caption{\label{fig:fig3} (a) Scheme of time-resolved measurements of the 4-wire voltage drop $V(t)$ across the junction at constant current bias $I_B$ vs DC (black dotted line) + AC (square-wave green line) gate voltage. The gate polarization signal is compared with the $I_C$ vs $V_G$ curve to highlight the dynamic working range of the system. (b) Voltage $V$ vs current $I$ characteristics for selected values of gate voltage $V_G$. The dot pairs show the working points of the system for different values of the bias current $I_B= 18,\ 37,\ 71\ \mu$A. 
(c) Evolution of the gate voltage vs time. The signal was obtained by summing a DC voltage $V_{DC}=10$ V and an AC square-wave voltage with amplitude $V_{AC}=5$ V. (d) Time-resolved voltage drop across the DB for selected values of the current bias. Dashed lines shows the corresponding working points in panel (b). These measurements were performed at $T_B=3$ K.}
\end{figure*}

Let us now describe further the transport properties of the transistor by focusing on the behavior of the device resistance $R$ under the action of the electric field. 
The steep modulation of the critical current yields, at constant current bias $I_B$,  an analogous modulation of the resistance of the device. Figure \ref{fig:fig2}(b) displays the evolution of $R$ as a function of the gate voltage $V_G$ for several values of bias current $I_B$. 
We note that $R$ remains almost equal to 0 for $V_G$ values such that $I_C(V_G)>I_B$. Then, for higher gate voltage, the resistance jumps to finite values due to the gate-driven superconducting-to-normal state transition of the device. Finally, $R$ increases with $V_G$, probably owing to the enlargement of the normal-state region within the junction \cite{Paolucci2018}. In particular, for $I_B>30$ $\mu $A, $R$ exhibits a linear dependence on $V_G$. 
Moreover, also the steepness of the resistance jump is a function of $I_B$, as customarily described by the figure of merit of the resistance derivative with respect to the gate voltage ($dR/dV_G$), which is shown in Fig. \ref{fig:fig2}(c). The amplitude of the differential resistance peaks, corresponding to the transition to the normal state, increases with $I_B$ reaching the maximum value of $\sim30$ $\Omega/$V at $I_B=150$ $\mu$A. Furthermore, the peak center shifts towards lower $V_G$ values as the bias current is increased.

The strong non linearity typical of the $R(V_G)$ characteristics, provided by the transition to the dissipative regime, could be conveniently exploited to implement a superconducting \emph{half-wave} rectifier  converting an AC signal applied to the gate into a DC one across the DB. 
Such a mechanism, recently proposed  for Nb gate-controlled transistors \cite{DeSimoni2020}, could allow to realize, for instance, a radiation detector in which the AC electric field collected by an antenna coupled to the gate electrode would be  rectified, thereby leading to an amplified non-zero average signal. 
The gain of such a device can be defined as the ratio $g=V_{out}/V_{in}$, where $V_{in}$ and $V_{out}$ are the peak-to-peak amplitudes of the signal applied to the gate and the voltage drop across the DB, respectively. 
The quantity $g$ can be calculated through the relation $g=R(V_G) I_B t/\sigma$ where, $R(V_G) I_B=V_{out}$ is the product of the gate dependent DB resistance $R$ and the current bias $I_B$, while $\sigma/t=V_{in}$ is given by the ratio of the typical width ($\sigma$) of the switching current probability distribution for gate-driven superconducting field-effect transistors \cite{Puglia2020} and
the transconductance $t=dI_C/dV_G$ \cite{Paolucci2019a}. 
For V-based devices, $g$ is expected to reach values as high as $\sim7$  with a typical power dissipation of  $\sim 40$ nW. 
Our V DB nanotransistor shows gain performance similar to cryogenic semiconductor devices \cite{Ivanov2011,Oukhanski2003} with a reduction in power consumption of about \emph{three} orders of magnitude. 
Noteworthy, a series of $N$ rectifiers can be realized by feeding the gate electrode of the $n^{th}$ rectifier with the output voltage of the $(n-1)^{th}$ one, and resulting in a total gain equal to $g^N$.

In the following, we shall focus on the time-resolved investigation of the response of V Josephson transistors to AC square-wave and sinusoidal gate signals. 
The measurements setup, comprising an input/output analog-to-digital/digital-to-analog converter (ADC/DAC) board for the generation and the acquisition of time-resolved alternate current and voltage signal, is depicted in Fig. \ref{fig:fig3}(a). The voltage drop $V(t)$ across the current-biased DB  was measured as a function of a time dependent gate voltage $V_G(t)$. 
In this configuration, when $I_C[V_G(t)]<I_B$ the junction is in the \emph{normal} state and a finite voltage drop is built across the DB (\textit{high-state}). By contrast, when $I_C(V_G(t))>I_B$, the DB is \emph{superconducting}, and $V\sim 0$ (\textit{low-state}). 
The response of the device to a transistor-transistor-logic (TTL) like square-wave excitation was probed by feeding the gate with a signal consisting of a $V_{DC}=10$ V DC bias added to a $V_{AC}=5$ V square-wave signal with frequencies up to $\sim50$ Hz, as shown in Fig. \ref{fig:fig3}(c). 
The voltage drop in the \textit{high-} and \textit{low-state} for some representative values of $I_B$ is shown in Fig. \ref{fig:fig3}(d) as pairs of points with the same color, superimposed on the $I-V$ characteristics taken at selected values of $V_G$. 
As displayed in Fig. \ref{fig:fig3}(d), the voltage drop across the junction preserves the shape of the input signal. 
On this regard, we emphasize that in our experiments the maximum operation frequency is limited by the low-pass filtering stages of the electrical lines of our setup. 
At present time, the timescale of gate-driven superconducting phase transition is  unknown, nonetheless the ultimate cut-off frequency of the V gate-controllable rectifiers could be estimated to be at most  $2\Delta / h\sim 265$ GHz, where $\Delta$ is the superconducting gap energy and $h$ the Planck's constant, a value which is comparable to state-of-art hybrid and CMOS field-effect transistors \cite{Valizadeh2016,Yeh2019,Horowitz1990}. 
\begin{figure}[t]
\includegraphics[width=\linewidth]{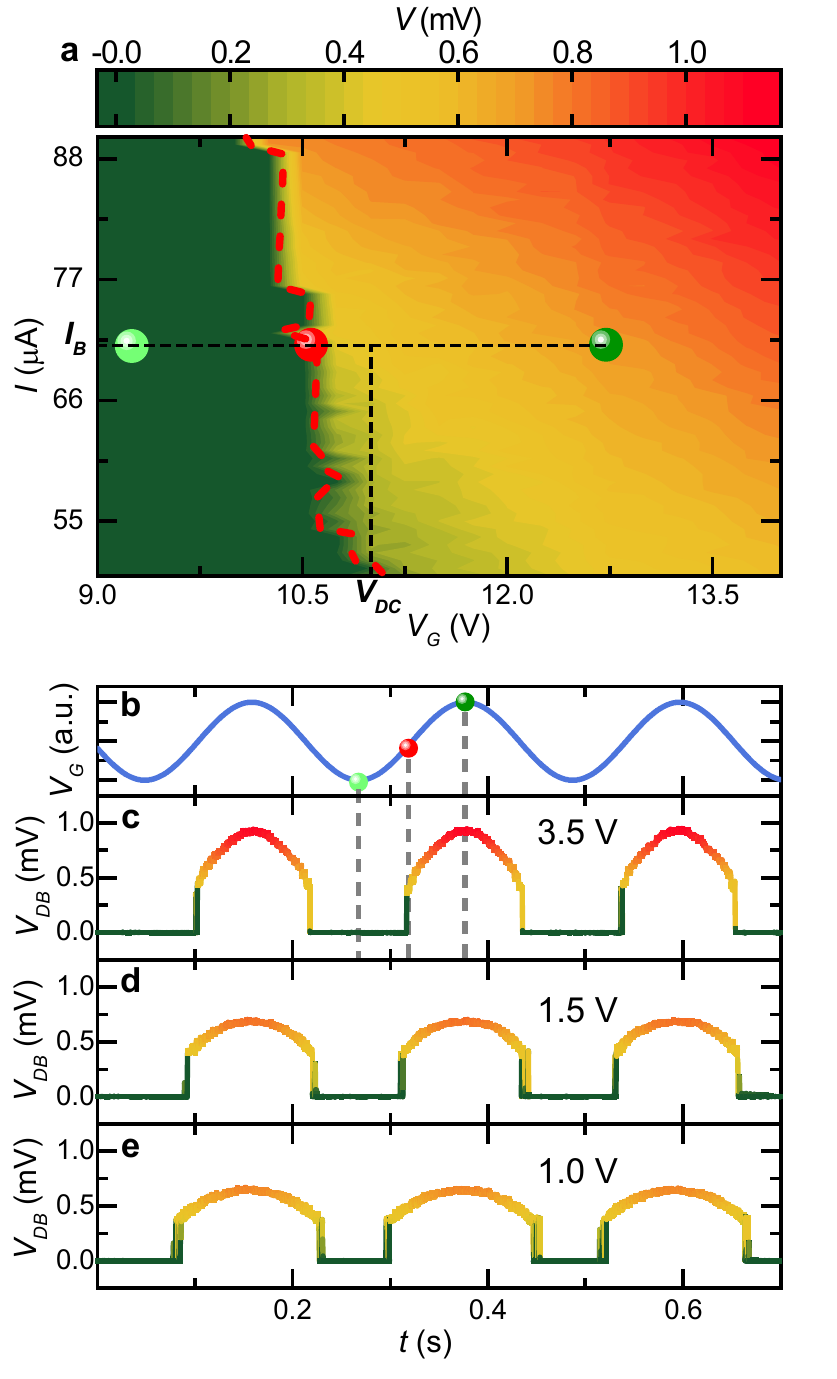}
\caption{\label{fig:fig4} (a). Density plot of the voltage drop across the junction as a function of the gate voltage ($x$ axis) and current bias ($y$ axis). From left to right, the three points represent the minimum zero resistance gate voltage value (light green), the transition between the superconducting and the normal state (red) and the maximum of both gate voltage and voltage drop (dark green) respectively. 
The dashed red curve highlights the critical current $I_C$ as a function of the gate voltage $V_G$. (b) Time evolution of the gate voltage.
The signal was obtained adding a DC voltage $V_{DC}=11$ V and an AC sine wave voltage $V_{AC}$. The three dots highlight the working points described in panel (a). (c), (d), (e) Time resolved voltage drop across the DB for, respectively, $V_{AC}=3.5,\ 1.5,\ 1.0$ V. The colormap is the same of panel (a). All these measurements were carried out at $T_B=3$ K.
}
\end{figure}

The response of the DB to a sinusoidal gate voltage excitation was measured by applying to the gate an AC sine signal $V_{AC}$ with peak-to-peak amplitudes ranging from 1 to 3.5 V added to a $V_{DC}=11$ V DC voltage bias. 
This results in a periodic sinusoidal oscillation of the critical current $I_C$ above and below $I_B$, producing a gate-driven periodic modulation of the device resistance. The voltage drop across the DB as a function of $I_B$ and $V_G$ is displayed in Fig. \ref{fig:fig4}(a). 
We note that by choosing $I_B$ and $V_{DC}$ [see Fig. \ref{fig:fig4}(a))], it is possible to set a working point for the rectifier, which determines the intensity and the shape of the resulting output voltage signal through the evolution of the super-to-normal switching point [red dashed line in Fig. \ref{fig:fig4}(a)], and through the dependence of the device resistance on the gate voltage (see also Ref. \cite{Paolucci2018}). 
Indeed, the system lies in the zero voltage state for $I_C(V_G(t))>I_B$, and a voltage drop across the junction occurs when $I_C(V_G(t))\simeq I_B$. For higher values of $V_G$, i.e., when $I_C(V_G(t))<I_B$, the voltage drop increases due to the gate-driven evolution of the DB resistance, which eventually will saturate at the asymptotic value of the normal-state resistance. 
Figure \ref{fig:fig4}(c) shows the voltage drop of the DB as a function of time, in response to the gate signal reported in Fig. \ref{fig:fig4}(b). 
To clarify the response of the system to such excitation, we highlight in Figs. \ref{fig:fig4}(a,b) three points corresponding to the minimum of the $V_G(t)$ signal (light green dot), the switching point (red dot), and the maximum of $V_G(t)$ (dark green dot). 
The value of the biasing current was set to $I_B=72\mu$A in order to take advantage of both the sharp normal-to-superconductor transition and the linear relation between $R$ and $V_G$ for $I_B>I_C$. 
The latter feature allows to preserve the sinusoidal shape in the output voltage when the JJ is in the resistive state, as shown in Figs. \ref{fig:fig4}(c,d,e) for different values of $V_{AC}$. Indeed, although the output signal resembles that of a conventional half-wave rectifier \cite{Horowitz1990}, it is worth to emphasize that, in our systems, it is possible to tune the rectification threshold $V_G^*$ by changing both $I_B$ and $V_{DC}$.

In summary, we have demonstrated a half-wave rectifier based on a gated  vanadium Dayem nano-bridge Josephson transistor operating at $3$ K. 
We showed the time-resolved response of our device to square-wave and sinusoidal gate signals, demonstrating the ability to convert an AC excitation into a DC one, and with a gain which is expected to obtain a value close to 7 at a power consumption of  $\sim40$ nW. 
The performance of the V DB represents a major improvement in energy efficiency compared to conventional cryogenic semiconductor technology \cite{Ivanov2011,Oukhanski2003}. 
V-based  gate-controllable  Josephson rectifiers could represent a breakthrough and a possible enabling technology to implement innovative all-metallic superconducting photon detectors \cite{Paolucci2020,Irwin1995,Irwin2006} based on gating as well as  electric diodes suitable for superconducting digital \cite{Barone1982} and quantum electronics \cite{Koch2007,Jazaeri2019} circuitry.

The  authors  acknowledge  the  European Union’s Horizon 2020 research and innovation program under grant No. 777222 ATTRACT (ProjectT-CONVERSE), and under grant No. 800923-SUPERTED for partial financial support.

\end{document}